\documentclass[sn-apa]{sn-jnl}

\usepackage{url}
\usepackage{hyperref}

\jyear{2023}
\begin{document}

\title[A Note on the Proposed Law for Improving the Transparency of Political Advertising in the European Union]{A Note on the Proposed Law for Improving the Transparency of Political Advertising in the European Union}

\author*[1]{\fnm{Jukka} \sur{Ruohonen}}\email{juanruo@utu.fi}
\affil*[1]{\orgdiv{Faculty of Technology}, \orgname{University of Turku},
\orgaddress{\street{Turun yliopisto}, \city{Turku}, \postcode{FI-20014}, \country{Finland}}}

\abstract{There is an increasing supply and demand for political advertising
  throughout the world. At the same time, societal threats, such as election
  interference by foreign governments and other bad actors, continues to be a
  pressing concern in many democracies. Furthermore, manipulation of electoral
  outcomes, whether by foreign or domestic forces, continues to be a concern of
  many citizens who are also worried about their fundamental rights. To these
  ends, the European Union (EU) has launched several initiatives for tackling
  the issues. A new regulation was proposed in 2020 also for improving the
  transparency of political advertising in the union. This short commentary
  reviews the regulation proposed and raises a few points about its limitations
  and potential impacts.}


\maketitle

\section*{Introduction}

The Cambridge Analytica scandal is the most well-known case about large-scale
computational interference in democratic elections and referendums. Although
debates continue about the scandal's actual impact upon historical election
outcomes, it is frequently raised as an alarming precedent by EU citizens
together with academics and many others~\citep{Brkan23, Hu20, Ruohonen23AIS,
  Zuboff22}. Although much of the controversy around the scandal was and
continues to be about politics in the United States, the waves hit also the
shores of Europe. Already before the Brexit vote, which too was allegedly a part
of the same scandal, it was reported that also political parties in the United
Kingdom had detailed databases on the political leanings of citizens in the
country~\citep{ORG19}. Thereafter, many smaller incidents have occurred
throughout Europe.

For instance, it was reported in 2020 that a data breach had occurred in Malta
that revealed the political opinions and other details of Maltese
voters~\citep{noyb20}. To put the actual breach aside, it remains unclear in
this case why such political data was collected in the first place and whether
the collection had any legal basis. Likewise, there is an ongoing case in
Austria involving its postal office's collection of personal data and using
algorithmic solutions for determining the political alignment of Austrians in
order to monetize the inferred data for political advertising
purposes~\citep{noyb22}. This case has also been escalated to the highest
European court.\footnote{~C-300/21.} In another recent case a civil society
group was unlawfully profiling people's political opinions and their other
details based on public social media data~\citep{Vanleeuw22}. Furthermore, not
only did political parties in the Hungarian 2022 general election use political
profiling and micro-targeting, but some of them also drew personal data
collected by the country's public administration for purposes other than
election campaigning \citep{HRW22}. Algorithmic political profiling has been
actively studied also in computer science (see \citealt{Baran22, Elmas23,
  Kitchener22, Qi23, Ram23}, among many others). The overall conclusion in these
academic studies is that a fairly good accuracy is achievable for determining
people's political stances based on public social media data. Obviously,
furthermore, social media companies and platforms have a much better vantage
point than academics to carry out political profiling about their users. It
should be remarked that in both cases European data protection laws
apply. Against the decisions on some of the noted recent cases, even individual
scientific researchers may be held legally liable for political and other
profiling. Ethical assessments are not enough.

In the aftermaths of the Cambridge Analytical scandal and related issues,
including the Russia's alleged interference with the elections in the United
States and other democratic countries, governments throughout the world
commenced studies on the topics of political profiling and election
interference. Regarding the European Union, the conclusion of many at the time
was that the legal competency and power were limited at the EU-level for
attempting to address the issues through regulative
action~\citep{Ruohonen20MISDOOM, LeinoSandberg19}. Against these assessments, it
is a small and positive surprise that the EU has managed to find new solutions
and reach a consensus over these. Although the General Data Protection
Regulation (GDPR) remains the cornerstone for tackling the issues, the recently
enacted Digital Services Act (DSA) is another notable recent large-scale
regulatory attempt in the European Union.\footnote{~Regulation (EU) 2016/679 and
  Regulation (EU) 2022/2065, respectively.} Proposals for improving transparency
of political advertising have been presented also under the the European
Democracy Action Plan.\footnote{~COM/2021/734 final.} Furthermore, the
non-binding Code of Practice on Disinformation was also strengthened in
2022. According to critics, however, this voluntary agreement together with
anything legally non-binding are doomed to fail~(see \citealt{Albert23, Avaaz21,
  Goujard23a}; and \citealt{Galantino23} for a recent academic assessment of the
criticism). Against this criticism, including the disappointment expressed by
some regarding the DSA~\citep{Killeen23, Ruohonen23AIS}, it is welcoming that
the European Commission launched in 2020 a further regulative initiative to
improve the transparency of political advertising and to restrict it to some
extent.\footnote{~COM/2021/731 final.} It is noteworthy that this proposal is a
regulation and not a directive; it is directly applicable EU law without the
necessity for national transpositions. Hence, it would patch noteworthy gaps in
national laws of some member states.

The European Parliament approved the proposal with a substantial majority in
early February 2023. Yet, at the time of writing, in October 2023, the proposal
is still fouling in the so-called trilogue negotiations. Therefore, it is a good
time to review the proposal and contribute to its deliberation with a few points
from a perspective of scientific research. By complementing the recent academic
assessment by \citet{Brkan23}, in what follows, the regulation proposed is thus
briefly reviewed after which a few points are raised about its potential impacts
and limitations.



\section*{The Proposal in Brief}

\subsection*{Motivation and Definitions}

The motivation for the proposal is specified in 71 non-enforceable recitals. The
first recital strikes at the heart of the problem: there has been an increasing
supply and demand for political advertising, which often occurs in a
cross-border manner through various actors, such as political consultancies,
advertising agencies and public relations firms, ad-tech platforms, political
data analytics firms~\citep{Simon19}, and so-called social media influencers who
too have recently expanded their controversial promotion activities toward
politics. Although the political activity of influencers has received a growing
interest in academic research~\citep{Arnesson22, DeGregorio22, Riedl21}, it is
arguably still the ad-tech industry together with platforms who are the main
facilitators or, depending on a viewpoint, the main culprits. This industry and
platforms continue to be also behind numerous data protection violations and
fundamental rights abuses according to EU citizens, civil society groups, and
media representatives; there are still threats even to democracy~\citep{Avaaz21,
  EPC21, Ruohonen23AIS}. This industry lacks also transparency as do
platforms. The same goes for online influencers who often fail to disclose their
commercial and political ties in a sufficiently transparent manner.

Therefore, the proposal notes in recital 4 that there is a legitimate public
interest to increase transparency in political advertising.\footnote{~On the
  legal grounds of Article~2 in the Treaty on European Union (TEU) as well as
  Articles~16 and 114 in the Treaty on the Functioning of the European Union
  (TFEU).} According to the proposal, such transparency is necessary to ensure a
healthy political debate and guarantee free and fair elections without
interference from bad actors, including those who spread disinformation through
advertising. Even though the freedom of expression is a fundamental right in the
EU, there are legitimate reasons to restrict it, as is the case with political
speech delivered through commercial advertising.\footnote{~On the legal grounds
  of Articles~11 and 52 in the Charter of Fundamental Rights of the European
  Union (CFREU), respectively.}  Then, in recital 5, the proposal continues to
discuss the platform-related issues in political advertising such as
micro-targeting and amplification. In recitals from 6 to 15 the proposal
discusses the lack of harmonization across the European Union, noting also that
the EU has an interest to promote pan-European political campaigning, including
campaigns done by the political groups in the European Parliament. Regarding
harmonization, there was some disagreements in the proposal's consultation: some
stressed that national laws are what matter~\citep{EACA21}, whereas others
argued that these national laws are woefully inadequate in some member
states~\citep{EPC21, PROVIDUS21}. Despite these disagreements, harmonization is
explicitly spelled out in the proposal's Article~3, which includes also a clause
for the member states who are forbidden to maintain existing laws or introduce
new laws that would diverge from the proposed EU regulation.

Furthermore and somewhat alarmingly from a data protection, fundamental rights,
and democracy perspective, the proposal is accompanied with many points about
innovation potential for political advertising in the context of the internal
market. Here, it remains debatable whether new technologies for political
advertising foster or hamper democracy, given the intrusive nature of online
advertising in general. If the EU politicians are truly concerned about the
ever-lasting democracy deficit in the union, there are far better means to
address the issue than innovation in political advertising. There seems to be
also some implicit plans to exploit the regulation for economic purposes, given
that Article~7(7) talks about some ``\textit{specific}'' but still entirely
undefined needs of micro, small, and medium-sized enterprises.\footnote{~As
  defined in Article 3 of Directive 2013/34/EU.} Given these points, it is also
noteworthy that the European Data Protection Board (EDPB) did not submit a
response to the open consultation that was held for the proposal. On the other
hand, the European Data Protection Supervisor, \citet{EDPS22}, has expressed an
opinion, arguing for a full ban of micro-targeting, stricter enforcement of the
GDPR's category of sensitive personal data, and prohibition of targeted
advertising based on pervasive tracking. Earlier, \citet{EDPB19} has stressed
that political profiling generally requires an explicit consent from data
subjects because political opinions belong to the GDPR's category of sensitive
data. As will be elaborated, also the proposal builds upon consent.

The definitions are specified in Article 2. To begin with, political advertising
refers to ``\textit{the preparation, placement, promotion, publication or
  dissemination, by any means, of a message by, for or on behalf of a political
  actor}'', excluding (a) purely private and commercial messages but (b)
including those ``\textit{liable to influence the outcome of an election or
  referendum, a legislative or regulatory process or voting behaviour}.'' There
are three relevant points about this definition. First, the lobbying from
advertiser interest groups such as \citet{IAB22} was taken into account; purely
commercial messages are excluded through the case~(a), although some media
representatives expressed a concern that the wording is not strong
enough~\citep{NME22}. In any case, these purely commercial ads should be
enforced through the GDPR. The second point is the confusing wording about
liability; in the EU jurisprudence this word connotes with legal liability,
which is not the case here~\citep{NME22, SNV22}. Third, it is precisely also
this case (b) that entails the difficult question over whether so-called ``issue
ads'' are covered; these are those ads that are used to bring public awareness
to certain public policy matters. Numerous respondents to the consultation
raised this problem, mostly arguing that these issue ads should not be covered
\citep{Google21, EACA21, IAB22}. While some argued that these are important when
advertising on noble policy matters such as waste reduction, the climate change,
or wildlife protection~\citep{NME22}, it can be equally argued that issue ads
could as well be about election fraud, xenophobia, healthcare disinformation,
and so forth. For instance, some platforms have allowed issue ads that have
contained death threats to election officers~\citep{C4D22}. Against this
backdrop, it is the duty of politicians, not industry associations or civil
society groups, to draw the proper line. If platforms and ad-tech are allowed to
make the choice, they likely continue to classify only a very small minority of
ads as political, often misclassifying political ads as
non-political~\citep{LePochat22, Ranganathan23}. Recital 17 tries to clarify
this issue, but fails to say anything relevant, including about a question of
who would have the power to determine whether a political advertisement is
``\textit{liable to influence}'' a policy outcome.

Then, political actors are defined to include (a)~national and European
political parties, (b)~political alliances, (c)~political candidates for posts
at European, national, regional, and local levels, (d)~already elected officials
at such levels, (e)~unelected members at such levels, (f)~political campaigning
organizations with or without legal personality, and (g)~any natural or legal
person representing or acting on behalf of the previously listed political
actors. Although some industry lobby groups stated that this kind of enumeration
of actors is worthless~\citep{EACA21}, it can be argued that this definition
captures well all the essential actors in European parliamentary politics. Of
these, the case (g) is noteworthy; it implies that also supporting associations
are covered, including labor unions, employer associations, business lobby
groups, civil society groups, and others insofar as they represent or act on
behalf of European political actors. In this regard, the wording in Article
2(4)(h) about ``\textit{any natural or legal person representing or acting on
  behalf}'' is somewhat difficult to interpret; it seems that lobbies who
advertise on policy matters in their own capacity are excluded. Furthermore, a
major limitation with the Article~(2)(4)'s definitions is that foreign forces
are excluded (unless they are natural or legal persons within Europe
``\textit{representing or acting on behalf}'' of European political actors and
``\textit{promoting the political objectives}'' of these). Hence, it remains
unclear whether the proposal is able to tackle the issue of foreign bad actors
buying political advertisements. As is well-known, this tactic has been a part
of election interference arsenal of Russia and other countries.

Because natural persons are covered, also individual citizens placing political
ads on behalf of politicians or political actors in general are in the
proposal's scope; an issue that was previously seen by some to be
problematic~\citep{Ruohonen20MISDOOM}. Also influencers are covered on the same
grounds. But, as clarified in recitals 27 and 29, this coverage applies only in
case commercial transactions are present; political content of individuals
posted in their purely personal capacity is excluded. In other words, citizens
and politicians are still free to promote organic political content. Given the
fears expressed by some that also non-commercial political speech might be
covered~\citep{Barata23, vanDrunen23}, these clarifications should be arguably
moved from the recitals to the articles. Finally, Article 2 defines also the
important concepts of targeting and amplification. These refer to
``\textit{techniques that are used either to address a tailored political
  advertisement only to a specific person or group of persons or to increase the
  circulation, reach or visibility of a political advertisement.}'' Hence,
targeting is about individual (micro-level) or group-based personalization.

\subsection*{Transparency Requirements}

Articles 5 to 11 and 14 specify the transparency requirements. In general, these
resemble those already specified in the DSA. Among other things, providers of
political ads are mandated to retain information on political campaigns, any
services provided for political advertising, the amounts of money invoiced and
the value of any other benefits offered, and, when applicable, the identify of a
sponsor and its contact details (Article 6). A sponsor is defined in Article~2
to be a ``\textit{natural or legal person on whose behalf a political
  advertisement is prepared, placed, published or disseminated}.''  Regarding
individual political ads, according to Article~7(1), these should contain
information in a clear, salient, and unambiguous manner about the fact that
these are political ads. Also the identity of a sponsor who placed given
political ads should be disclosed and there should be a specific transparency
notice for each political ad.

Article 7(2) then specifies the requirements for the transparency notices. Among
these are: the identify of a sponsor and its contact details, the period a given
ad is shown, aggregated data about money spent for an ad, information about
particular elections or referendums a non-issue ad seeks to impact, and links to
ad repositories. Regarding the last two of these mandates, Articles 7(2)(d) and
7(2)(e) both prefix their mandates with a wording about ``\textit{where
  applicable}''. This wording leaves it unclear whether all political
advertising services must maintain repositories, although very large platforms
are mandated to do so according to Article~7(6) and its references to the
DSA. This mandate will hopefully solve some of the well-known data access
problems encountered by researchers~\citep{Galantino23}. Machine-readability is
promoted too, which is a welcome addition because the repositories of some
companies have not provided such functionality~\citep{Ranganathan23,
  Ruohonen20MISDOOM}. Though, Article~7(4) makes an undesirable relaxation by
using a wording ``\textit{where technically possible}''. From a computer science
perspective, there should be no technical obstacles whatsoever for providing
machine-readability.

Then, in Article~8, political advertising publishers (which are defined in
Article~2 to be natural or legal persons who broadcast, make available through
an interface, or otherwise bring political ads to the public fora through any
medium) are required to disclose money or other benefits received in their
annual financial statements.\footnote{~As specified in Article 19 of Directive
  2013/34/EU.} This disclosure statement was seen as excessive by some
stakeholders~\citep{NME22}. According to Article~9, political advertising
services and publishers must also put a mechanism in place to allow any
individual to notify about political ads that infringe the proposed regulation.

Articles 10 and 11 specify transmission requirements for political ads to
competent authorities and other interested entities. Regarding the competent
public authorities, it is not entirely clear to whom these refer since election
laws vary greatly across the member states. Although some EU-level guidance
exists, recital 57 and Article 15(2) specify that the member states should
designate new competent authorities for this task in conjunction with the
national coordinators specified in the DSA.\footnote{~Recommendation C(2018)
  5949 final.} These new competent authorities are also granted a power to
impose fines according to Article 15(5)(c). As argued by \citet{Kozak23}, and
analogously to the DSA~\citep{Ruohonen23AIS}, the \textit{de jure} independence
requirement for these authorities may be problematic in those member states
where \textit{de facto} dependencies may be present.

As for political profiling, targeting, and amplification, the competence
question is clearer: these are national and EU-level data protection
authorities, as clarified in the proposal's Article~15(1). Consequently, the
proposal brings a further issue about the enforcement of the GDPR toward
political parties and other political actors in Europe, including a question
about who has the power for this enforcement at the EU-level; here it is likely
the European Data Protection Supervisor~\citep{Brkan23}. Regarding the
interested entities, these are specified to be vetted researchers in line with
the DSA, civil society groups, political actors, and election observers
accredited in the member states, all of whom must be independent from commercial
interests. Even though charging fees is allowed by Article~11(5), some media
representatives argued that there should a right for political advertising
services to outright refuse transmission requests from these actors in case
their requests are unfounded~\citep{NME22}. Such a right would seriously hamper
accountability assessments, including those done in scientific research. There
are also some other concerns about the impacts upon scientific research; these
are discussed later on.

\subsection*{Political Profiling}

Article 12 is noteworthy. It prohibits both targeting and amplification of
political ads based on the GDPR's category of sensitive personal data. As
specified in the latter's Article~9, among other things, this category involves
``\textit{personal data revealing racial or ethnic origin, political opinions,
  religious or philosophical beliefs, or trade union membership}.'' This
prohibition applies also to personal data of people working in EU
institutions.\footnote{~As specified in Article~10 of Regulation (EU)
  2018/1725.} Though, there are two relaxations: targeting based on sensitive
personal data can be still done in case a person has given an explicit consent
for it or in case he or she is a member or a past member of a not-for-profit
body with a political, philosophical, religious, or trade union aim. These are
specified in the GDPR's Article 9(2)(a) and~(d). Besides these two exemptions,
the real issue here is that companies and others should not be collecting sensitive
personal data to begin with; the GDPR generally prohibits processing of such
data without a consent.

Though, regarding political profiling based on public sources, such as in the
noted academic computer science studies, a notable weakness is the exemption in
the GDPR's Article 9(2)(e) according to which processing is allowed in case a
person has manifestly made his or her sensitive personal data public. European
politicians and political parties might also try to justify political profiling
on the grounds that they maintain democracy and have legitimate interests for
electoral purposes, as noted in the GDPR's recital 56, and then using the GDPR's
exemption in Article 9(2)(g) on substantial public interests as the legal basis
together with Article~6~\citep{Brkan23}. The same point has been made by
\citet{EDPS18}. But since the European political actors rely on platforms,
ad-tech, political data analytics companies, and other third-parties for the
political profiling, targeting, and amplification, these kind of justifications
would arguably have only weak grounds. The same applies to civil society groups
who, following the arguments of some academics about their special but
mysterious get-out-of-jail card to political profiling and targeting based on
sensitive personal data~\citep{vanDrunen23}, might try to rely on the GDPR's
exemption in Article~9(2)(d)~\citep{Brkan23}. These exemptions in the GDPR bring
also a further question about the GDPR's Article~22 on the right to object
automated decision-making, including profiling. However, this is article is
generally seen problematic in many ways \citep{Nisevic22}, and there has been
little guidance from data protection authorities on this matter, particularly in
the context of political profiling. In essence, \citet[p.~26]{EDPB21} has only
noted that it is the duty of a social media company to determine whether the
right applies with respect to the Article's wording about significant effects.

These points were missed by some responding to the proposal's consultation who
argued that sensitive personal data can only be processed based on
consent~\citep{EACA21, NME22}. However, even under Article~9(2)(d),~(e),~(g), or
the journalistic exemption specified in the GDPR's Article~85, the recent
interpretation of national data protection authorities seems to be that many
other relevant conditions still apply, including those specified in the GDPR's
Articles~5, 12, and~14~\citep{Vanleeuw22}. Also Articles~24, 25, 28, 32, and 35
must be taken into account~\citep[p.~31]{EDPB21}. Of these, Article~25 is
typically violated by platforms engaging in micro-targeting; they do not obey
the principle of data protection by design and by default~\citep{Casagran21}. In
any case, according to the proposed law, actual targeting of political
advertisements based on sensitive personal data is still prohibited unless the
two noted exemptions apply. If these do apply, Article~12 in the proposal then
mandates that all targeting and amplification techniques and their parameters
should be described in plain English in internal policy documents. Explanations
to laypeople about these should be specified in each political ad. Here, it
remains generally unclear what exact information should be disclosed alongside
political ads. It could be argued that the relevant information about techniques
and parameters contains all the personal data a platform holds about a
person. This kind of interpretation would imply that the transparency mandate
reduces to the right to access personal data granted by the GDPR.

More generally, many respondents to the consultation emphasized that the
foremost issue is a proper enforcement of the GDPR, as has been stressed also by
academics, EU citizens, and many others \citep{EDPS22, NME22, NOYB23a,
  Ruohonen22IS, Ruohonen23AIS}. Instead of attempting to solve the issues
through transparency or ``\textit{the curtailment of the granularity of
  targeting possibilities in the context of political messaging, perhaps a
  similar result could be reached through strict enforcement of data protection
  rules with respect to political campaigns in Europe}''
\citep[p.~35]{vanHoboken21}. In this regard, it is a surprise that even
\citet{IAB22} was strongly in favor of enforcing the GDPR's Article 9. The same
point was made by some think-tanks with an alignment toward civil society who
agreed but further recommended that the use of ``\textit{inferred data should be
  banned without exceptions}'' particularly in the context of political
advertising~\citep{SNV22}. Similar points about inferred data were raised during
the consultation for the DSA, although also these were bypassed by
lawmakers. But regarding the GDPR, the European Commission is planning to
address the many problems with the regulation's
enforcement~\citep{Goujard23b}. A~more active role of the Commission is also
envisioned in the proposed law for political ads.\footnote{~On the legal grounds
  of Article 290 in the TFEU.} According to recital 66 and Article 19, the
Commission is powered to adapt delegated acts for the transparency requirements
in Article 7 together with the profiling and amplification issues specified in
Article 12. This provision is welcome because specifying some vague details
about ``main parameters'' of machine-learning algorithms is not sufficient to
improve transparency, as has been pointed out also with respect to the
DSA~\citep{Ruohonen23AIS}. In fact, it is arguably even impossible to specify
such parameters in newer algorithms; there are reportedly as many as 175 billion
parameters in the celebrated ChatGPT.

\section*{Conclusion}

This short commentary reviewed the law proposal to improve transparency of
political advertising in Europe and to restrict targeting and amplification
based on political profiling. The proposal is generally a welcome addition to
the EU's toolbox for tackling societal threats, including those for
democracy. Six points can be made about the proposal's potential impact and
consequences.

First, the proposal is somewhat unclear regarding its impact upon fundamental
rights. It is claimed that fundamental rights are improved regarding the CFREU's
Article 8, that is, the right to the protection of personal data, while minimal
restrictions are placed upon the freedom of expression and information
(Article~11). Of these, the claims about Article~8 are sufficiently clarified,
but the explanations regarding the impact upon Article~11 remain vague. It is
also noted that the right to private life (Article~7) is affected, but nothing
is said beyond mentioning this right. Given that the Court of Justice of the
European Union has not thus far managed to build sufficient case law regarding
data-driven electoral campaigning~\citep{Brkan20}, some vagueness regarding
fundamental rights likely persists also in a future law based on the proposal.

Second, it remains debatable whether the proposed regulation will be able to
curtain micro-targeting because it still allows political profiling and
targeting based on the consent of citizens. As has been frequently argued,
people neither understand to what they are consenting to nor are sufficiently
aware that they are being profiled after their consents have been given
\citep{Brkan23, Casagran21, Ruohonen23AIS, Ruohonen23DS}. It can be argued that
it is specifically those people who do not understand the situation with their
consents who are also the most prone to be influenced by micro-targeting. The
reliance on consent continues to be a major issue in the whole European data
protection regime. It can be also argued that micro-targeting is not the real
danger; data and platforms can be weaponized against democracies by using many
ways of manipulation, including interference in the democratic deliberation
process during electoral campaigning as well as in-between elections. While
disinformation and cyber attacks are the prime examples~\citep{LeinoSandberg19},
advances in generative artificial intelligence prompt a speculation about
whether the whole public fora will be flooded with ``manipulated'', inauthentic,
machine-generated content.

Third, the regulation proposed does not address some of the long-standing issues
in political advertising. Although transparency requirements are imposed,
nothing is mandated for campaigning finances and their upper limits. Thus, the
proposal does not solve the problem of some particular political parties and
other political actors having ample and unfair financial resources for political
advertising~\citep{Kozak23}. Even though the electoral campaigning budgets in
Europe are far from those in the United States, the amount of money spent on
lobbying at Brussels is enough to warrant a concern---especially when
considering that the proposal's scope covers all political messages targeted to
influence any given policy outcome.

Fourth, many stakeholders who responded to the proposal's consultation
emphasized a need to establish a cross-platform and real-time political
advertisement repository~\citep{PROVIDUS21, SNV22}. Such a repository would be
highly valuable also for scientific research. The best available solution for
science would be a unified repository maintained by the EU itself. Such a
repository would allow to innovate in applications that foster democracy,
accountability, and transparency. In this regard, it is a disappointment that
the proposal outsources the repository question to political advertising
publishers, services, and platforms. According to Article~7(5), political
advertising publishers must retain data about their transparency notices for a
period of five years. This period is far too short from a perspective of
scientific research. Fields such as political science, sociology, and history
are slow-paced and they often address important historical questions. It could
be even argued that online political ads belong to the Europe's cultural
heritage. Someone writing about the history of politics in the EU in, say, 2040,
might find online political ads as relevant for better understanding the
long-run political developments and democracy in the union. With advances in
artificial intelligence, the so-called digital humanism is likely to advance as
well. Given that there are legitimate interests for political advertising
publishers in this regard, such as the economic costs related to retaining
political advertisement data, these points restate that a better solution would
be a unified database maintained by a competent EU institution. Given that
platforms have been seen as biased toward collaborating with universities in the
United States and particularly the Ivy League institutions
therein~\citep{Dommett22}, a unified EU-level solution would also offer
opportunities for European research institutions.

Fifth, the proposal as well as the DSA raise a more general concern about
scientific research and scholarship. It is well-understood that also scientists
and researchers must be vetted; after all, the whole Cambridge Analytica scandal
was related to ethical lapses of scientific researchers. That said, a concern
remains about the vetting in terms of potential discrimination. Competition in
science is fierce and the so-called Matthew effect \citep{Merton68} has only
increased. Thus, there are legitimate arguments that poorer universities and
those researchers who fail to win grants might be unfairly left behind or even
prevented from having access to relevant data. There are also well-known
differences between disciplines; humanities and social sciences are in most
countries under a more severe financial pressure compared to the hard
sciences. This point applies also to the EU's science funding such as the
large-scale Horizon Europe. Yet, it is precisely humanities and social sciences
who would most benefit from data access to political advertising and other
related data repositories. That is to say: not only may the vetting affect
inequalities between universities but it may also affect inequalities between
disciplines. A similar point can be raised regarding applied research done by
civil society groups: the vetting may benefit those particular groups who are
perceived favorably by national governments and the EU who also often fund the
groups. Smaller national groups might be left behind and the Brussels-based ones
might receive an unfair advantage. There is also an argument to be made about a
need to strengthen the often smaller civil society groups in the member states
within which democratic backsliding is occurring.

Sixth, the regulation may entail a risk that political parties shift their
advertising budgets toward the creation of their own non-regulated organic
content, which would render the regulation meaningless and affect the economic
situation of publishers~\citep{NME22}. In other words, disagreements exist over
whether political campaigning is better done through advertising or through
organic content. An anecdotal impression from the Finnish 2023 parliamentary
election campaigning was that politicians indeed increasingly turned toward
promoting their organic political content. Though, this promotion seemed to
still occur mainly on the large global social media platforms, including also
those with alleged ties to intelligence and military agencies in authoritarian
regimes. From a citizen perspective, no real improvements have occurred;
following politics, whether national or European, often still implies
surrendering fundamental rights to the mercy of platforms and whoever is behind
the scenes. The same point extends toward public administrations who too
increasingly rely on foreign platforms for delivering their informing
obligations. Here, there is room for the EU to innovate in developing its own
platforms on which European citizens, politicians, civil servants, businesses,
and others could safely discuss issues without a fear of fundamental rights
abuses and manipulation by bad actors, including those sponsored by hostile
foreign governments. This argument finds its counterpart in fundamental
rights. As recently argued by \citet{Giglio23}, a key element in the fight
against disinformation is actually explicitly spelled out in the CFREU's Article
11, but not necessarily in a sense that more ``good'' speech would solve the
situation, but in that this fundamental right contains the often overlooked
right ``\textit{to receive and impart information and ideas without
  interference}''. At the moment, it remains unclear whether European public
administrations, including those of the European Union, fulfill this fundamental
right of citizens due to their increasing reliance on platforms.

\section*{Epilogue}

As it seems, there are currently two obstacles in the
negotiations~\citep{EDRI23a}, both of which were also addressed in this short
note. The first originates from the fears that also non-commercial political
speech might be covered. The second is about the use of special categories of
personal data in political advertising, and whether a consent of citizens
suffices for the use. Although both are thorny issues, the data protection
concerns have progressed outside of the EU's negotiation rooms. For instance,
complaints have already been filed against political parties for their
micro-targeted political advertisements on social media
platforms~\citep{noyb23b}. Given the enduring problems with the GDPR's
enforcement, it may be that also these complaints will end up to
courts. Enforcement seems to be looming also with the DSA. The
\citeauthor{EC23a}'s \citeyearpar{EC23a} preliminary investigations indicate
that some social media platforms have been either unable or unwilling to tackle
the flood of Russian disinformation in the face of the war in Ukraine. In the
meanwhile, the Commission itself has used political micro-targeting in an
attempt to influence public opinion on another highly controversial law
proposal, prompting some observers to accuse the Commission of the very same
shady disinformation tactics used by Russia in the past~\citep{Mekic23}. So
controversial has the Commission's behavior been that the executive branch of
the European Union may now be a subject of an investigation conducted by the
EDPS~\citep{Tar23a}. All animals are equal but some animals are more equal than
others.

\section*{Acknowledgements}

The author thanks Kalle Hjerppe for comments.


\end{document}